\documentclass[prl,twocolumn,showpacs]{revtex4}  % for review and submission
\usepackage{graphicx}  % needed for figures
\newcommand{\Cosh}{ch}
\newcommand{\Sinh}{sh}
\newcommand{\Tanh}{th}
\newcommand{\Ash}{arcsh}

\begin{document}

% the following line is for submission, including submission to the arXiv!!
%\hspace{5.2in} \mbox{Fermilab-Pub-04/xxx-E}

\title{Two-elliptic coordinates. Eigenfunctions and eigenvalues.}

%\input author_list.tex       % D0 authors (remove the first 3 lines
                             % of this file prior to submission, they
                             % contain a time stamp for the authorlist)
                             % (includes institutions and visitors)
\author{G.\,V.\,Kovalev \/\thanks
{e-mail: kovalevgennady@qwest.net}}

%%% author(s) - for colontitle (at the top of the page)
%\rauthor{G.\,V.\,Kovalev}

%%% author(s) - for table of contents
%\sodauthor{Kovalev}

%%% author's address(es)
\address{North Saint Paul,  MN 55109, USA }

%%% dates of submition & resubmition (if submitted once, second argument is *)
%\dates{10 November 2013}

\date{\today}

\begin{abstract}
The elliptic coordinates are used to build a new families of 2D coordinate systems which are orthogonal and admits the separation of variables. The charts of characteristic curves  are constructed for these systems and compared with Mathieu's. The possible applications are in quantum mechanics, diffraction theory, vibrations and classical dynamics.
\end{abstract}

%%% PACS numbers
\pacs{03.65.Ge; 03.65.Fd; 03.65.׷; 03.65.Db;  02.30.Em}
\maketitle

%\section{\label{sec:level1}First-level heading}
% sections are not used for PRL papers

\textbf{I.Introduction.}  It is well known  that main source of orthogonal coordinate systems (CS) is a conformal transformation, but the use of analytical functions for separation of Schr\"{o}dinger (SE) or  Helmholtz (HE) equations in 2D space is restricted to only 4 CS: Cartesian, polar, parabolic and elliptic \cite{MorseFeshbach:1953, Miller:1976}. The first 3 CS are  unique and can be considered as degenerated cases of the last one. In fact, the elliptic CS includes an infinite number of CS, each of which is generated by one focal distance $f$. The combination of several elliptic CS with different foci in one CS \cite{Kova:2013} is a new unlimited source of orthogonal coordinate systems which can admit  the separation of variables for SE and HE. Several single CS with Cartesian and polar CS have been already in use for a long time (see e.g. the ``stadium billiard'' \cite{Bun:1974}), but the approach with several elliptic CS matching each other by special rules can satisfy much larger variety of shapes  and boundaries for the purpose of separation. 

\begin{figure}
	\centering
		\includegraphics[width=0.50\textwidth]{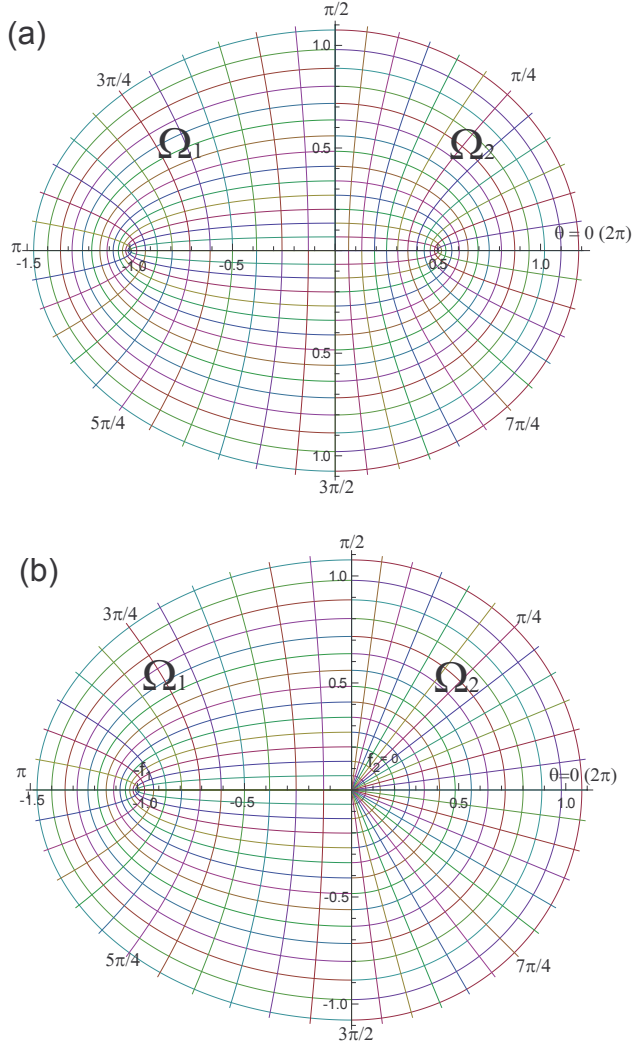}
	\caption{(a)Two-elleptic CS with foci $f_1=1$,  $f_2=0.5$.  (b) Degenerated case with foci $f_1=1$, $f_2=0$.	}
	\label{fig:Two-elliptic}
\end{figure}

\textbf{II. Separated ODE and solutions.}
The simplest CS is composed of two elliptic CS with different foci $f_1$, $f_2=\alpha f_1$ ($\alpha$ is a scale coefficient) Fig.\ref{fig:Two-elliptic}(a).  This two-elliptic CS \cite{Kova:2013} with only symmetry along $x$ axis is  
\begin{eqnarray}
x = f_1 \left \{\begin{array}{ll} \alpha \Cosh[ \Ash (\frac{\Sinh \mu }{\alpha} )] \cos \theta, &  \;  -\frac{\pi}{2} \leq \theta \leq \frac{\pi}{2}, \\  
\Cosh \mu \cos \theta,&  \;  \frac{\pi}{2} \leq \theta \leq \frac{3\pi}{2},
\end{array} \right. \nonumber \\
y = f_1\left \{\begin{array}{ll} \alpha \Sinh[ \Ash (\frac{\Sinh \mu }{\alpha})] \sin \theta, &  \;  -\frac{\pi}{2} \leq \theta \leq \frac{\pi}{2}, \\  
\Sinh \mu \sin \theta,&  \;  \frac{\pi}{2} \leq \theta \leq \frac{3\pi}{2},
\end{array} \right.
\label{eqw1}
\end{eqnarray}
and the Helmholtz equation separates $\Psi(\theta,\mu)=\Phi(\theta) R(\mu)$ leading to two ODE for $\Phi(\theta)$and $R(\mu)$:
\begin{eqnarray}
\Phi^{''} -\frac{g_1^{'}}{2g_1}\Phi^{'} +g_1[\lambda + k^2 h_1]\Phi=0,\label{eqw2a} \\ 
R^{''} -\frac{g_2^{'}}{2g_2}R^{'}-g_2[\lambda -  k^2 h_2]R=0.
\label{eqw2b}
\end{eqnarray}
Here $\lambda$ is the constant of separation, $g_1=1$ and  $g_2,h_1,h_2, g^{'}_2/2g_2$ are  piecewise continuous functions of $\theta$ or $\mu$ only:
\begin{eqnarray}
h_{1} =-\frac{ f_1^2 \cos 2 \theta}{2} \left \{\begin{array}{ll} \alpha^2, &  \;  -\frac{\pi}{2} \leq \theta \leq \frac{\pi}{2}, \\  
1&  \;  \frac{\pi}{2} \leq \theta \leq \frac{3\pi}{2}.
\end{array} \right.\nonumber \\ 
h_{2} = \frac{f_1^2}{2} \left \{\begin{array}{ll}  \Cosh 2\mu +\alpha^2-1 , &  \;  -\frac{\pi}{2} \leq \theta \leq \frac{\pi}{2}, \\  
  \Cosh 2\mu &  \;  \frac{\pi}{2} \leq \theta \leq \frac{3\pi}{2}.
\end{array} \right. \nonumber \\
g_2 = \left \{\begin{array}{ll} \frac{\Cosh^2 \mu }{\alpha^2+ \Sinh^2 \mu}, &  \;  -\frac{\pi}{2} \leq \theta \leq \frac{\pi}{2}, \\  
1,&  \;  \frac{\pi}{2} \leq \theta \leq \frac{3\pi}{2},
\end{array} \right. \nonumber \\ 
\frac{g^{'}_2 }{2 g_2 }= \left \{\begin{array}{ll} \frac{(\alpha^2 -1) \Tanh \mu }{\alpha^2+ \Sinh^2 \mu}, &  \;  -\frac{\pi}{2} \leq \theta \leq \frac{\pi}{2}, \\  
0,&  \;  \frac{\pi}{2} \leq \theta \leq \frac{3\pi}{2}.
\end{array} \right. 
\label{eqw4}
\end{eqnarray}
Thus the Eq. \ref{eqw2a}, \ref{eqw2b}  for $ -\frac{\pi}{2} \leq \theta \leq \frac{\pi}{2}$ are:   
\begin{eqnarray}
\Phi^{''}  +[\lambda - 2 q_2 \cos 2 \theta]\Phi=0,\label{eqw5a} \\ 
R^{''} - \frac{(\alpha^2 -1) \Tanh \mu }{\alpha^2+ \Sinh^2 \mu}R^{'}- \nonumber \\
\frac{\Cosh^2 \mu }{\alpha^2+\Sinh^2 \mu}[\lambda -  2q_1(\Cosh 2\mu +\alpha^2-1)]R=0;
\label{eqw5b}
\end{eqnarray}
and for $ \frac{\pi}{2} \leq \theta \leq \frac{3\pi}{2}$ are:
\begin{eqnarray}
\Phi^{''} +[\lambda -2 q_1 \cos 2 \theta]\Phi=0, \label{eqw6a} \\
R^{''} -[\lambda - 2 q_1 \Cosh 2 \mu]R=0,
\label{eqw6b}
\end{eqnarray}
where we use the standard notations $2 q_1 =\frac{k^2  f_1^2}{2}$, $2 q_2 =\frac{k^2  f_2^2}{2}=\frac{k^2 \alpha^2 f_1^2}{2}=2 q_1 \alpha^2 $. 
 If the scale coefficient $\alpha = 1$, the Eqs. \ref{eqw5a}-\ref{eqw6b} become identical to Mathieu's. Otherwise, they are slightly different, and a small deviation $\alpha $ from 1, makes essential  changes in the spectrum behavior of the problem. 
In particular, the Mathieu's periodic condition $\Phi(0) = \Phi(\pi)$ does not hold anymore. Only   condition $\Phi(0) = \Phi(2 \pi)$ is supported in the angular equations  \ref{eqw5a},\ref{eqw6a}.

To find the eigenvalues and eigenfunctions in our CS for 2D free space, we write down the general solution in two regions, which are  expressed through fundamental set of  Mathieu's functions $ce_{\nu}(\theta, q)$, $se_{\nu}(\theta, q)$ in each region   
\begin{eqnarray}
 \Phi_{1} =A_1 ce_{\nu}(\theta,  q_1)  + B_1 se_{\nu}(\theta, q_1), \nonumber \\
 \Phi_{2} =A_2 ce_{\nu}(\theta, q_2)  + B_2 se_{\nu}(\theta, q_2),
\label{eqw7}
\end{eqnarray}
where $\nu$ is not an integer, in general.

Now we require the continuity and differentiability of the solution at the boundary $\theta=\pi/2$
\begin{eqnarray}
 \Phi_{1}(\frac{\pi}{2}) =\Phi_{2}(\frac{\pi}{2}), \nonumber \\
 \Phi_{1}^{'}(\frac{\pi}{2}) =\Phi_{2}^{'}(\frac{\pi}{2}),
\label{eqw8}
\end{eqnarray}
and compliancy with Bloch's theorem on the boundary $\theta=-\pi/2$ ($\theta=3\pi/2$): 
\begin{eqnarray}
 \Phi_{1}(\frac{3\pi}{2})e^{i 2 \pi K} =\Phi_{2}(-\frac{\pi}{2}), \nonumber \\
 \Phi_{1}^{'}(\frac{3\pi}{2})e^{i 2 \pi K} =\Phi_{2}^{'}(-\frac{\pi}{2}).
\label{eqw9}
\end{eqnarray}
(Note, that we distinguish characteristic exponent $K$ for two-elliptic CS from $\nu$ for Mathieu's functions.)
  
Substituting expressions Eq. \ref{eqw7}
in Eqs. \ref{eqw8}, \ref{eqw9}, we receive 4 Eqs. for unknown $A_1, B_1, A_2, B_2$. 
The  coefficients at $A_1, B_1, A_2, B_2$ constitute  $4 \times 4$ matrix $M$:
\begin{eqnarray}
\left[ \begin {array}{cccc} ce_{\nu}(\frac{\pi}{2},q_1)&se_{\nu}(\frac{\pi}{2},q_1)&-ce_{\nu}(\frac{\pi}{2},q_2) &-se_{\nu}(\frac{\pi}{2},q_2)\\\noalign{\medskip} ce'_{\nu}(\frac{\pi}{2},q_1)&se'_{\nu}(\frac{\pi}{2},q_1)&-ce'_{\nu}(\frac{\pi}{2},q_2) &-se'_{\nu}(\frac{\pi}{2},q_2)\\\noalign{\medskip} \beta ce_{\nu}(\frac{3\pi}{2},q_1)&\beta se_{\nu}(\frac{3\pi}{2},q_1)&-ce_{\nu}(-\frac{\pi}{2},q_2) &-se_{\nu}(-\frac{\pi}{2},q_2) 
\\\noalign{\medskip}\beta ce'_{\nu}(\frac{3\pi}{2},q_1)& \beta se'_{\nu}(\frac{3\pi}{2},q_1)&-ce'_{\nu}(-\frac{\pi}{2},q_2) &-se'_{\nu}(-\frac{\pi}{2},q_2)
  \end {array}
 \right].
\label{eqw10} 
\end{eqnarray} 
We use in \ref{eqw10} the notation: 
$\beta = \exp i 2 \pi K$.  
In order  to solve the system  \ref{eqw8},  \ref{eqw9}, the determinant of the matrix $M$ must satisfy the equation
\begin{eqnarray}
 \det(M)=0.
\label{eqw11}
\end{eqnarray}
This is, basically, the equation for eigenvalues $\lambda$ vs., say, $q_1$ if the ratio $\alpha=f_2/f_1$ is fixed. 
The solution of \ref{eqw11} can be written in the form
\begin{eqnarray}
 \cos 2 \pi K=\frac{F(\lambda,q_1,q_2)}{2 \det(W[q_1])\det(W[q_2])}.
\label{eqw12}
\end{eqnarray}
where $\det(W[q_1]),  \det(W[q_2])$ are Wronskians for fundamental solutions with $q_1$ or $q_2$
\begin{eqnarray}
 \det(W_{q})=ce'_{\nu}(,q)se_{\nu}(,q)-ce_{\nu}(,q)se'_{\nu}(,q),
\label{eqw13}
\end{eqnarray}
 (argument $\theta$ does not matter).
The function $F(\lambda,q_1,q_2)=F(\lambda,q_1,q_1 \alpha^2)$ is bulky expression from determinant calculations. Instead of $ce_{\nu}(\theta,q)$,  $ce'_{\nu}(\theta,q)$, $se_{\nu}(\theta,q)$, $se'_{\nu}(\theta,q)$ we introduce  the short notations: $c_{\theta,q},c'_{\theta,q},s_{\theta,q},s'_{\theta,q}$ with indices $(q_1,q_2)=(1,2)$, $(\frac{\pi}{2}, -\frac{\pi}{2},\frac{3\pi}{2})=(1,2,3)$. After that,  the expression $F(\lambda,q_1,q_1 \alpha^2)$ can be written down in the compact form:
\begin{eqnarray}
F(\lambda,q_1,q_1 \alpha^2)= & \nonumber \\
 c_{31}c'_{12} s_{12} s'_{11}+ c_{11}c'_{12}s_{12}s'_{31}-2c_{31}c'_{12}s_{11}s'_{12}+ \nonumber \\
2c_{11}c'_{12}s_{31}s'_{12}+c_{31}c_{12}s'_{11}s'_{12}+c_{11}c_{12}s'_{31}s'_{12}- \nonumber \\  
c'_{31}(c'_{12}s_{11}s_{12}+c_{12}(-2s_{12}s'_{11}+s_{11}s'_{12}))- \nonumber \\
c'_{11}(c'_{12}s_{31}s_{12}+c_{12}(2s_{12}s'_{31}+s_{31}s'_{12})).    
\label{eqw14} 
\end{eqnarray} 
The right side of Eq. \ref{eqw12} can not be out of the range $(-1,1)$ of the left side \ref{eqw12}, and two conditions:
\begin{eqnarray}
\frac{F(\lambda,q_1,q_1 \alpha^2)}{2 \det(W[q_1])\det(W[q_1 \alpha^2])}=1 \label{eqw15a}, \\
\frac{F(\lambda,q_1,q_1 \alpha^2)}{2 \det(W[q_1])\det(W[q_1 \alpha^2])}=-1,
\label{eqw15b}
\end{eqnarray}
define the eigenvalues vs. parameter $q_1$ (characteristic curves)  for periodic solutions of the system  \ref{eqw5a}, \ref{eqw6a}. These curves separate the stable solutions from unstable  and on these curves the characteristic exponent 
must be an integer number. 

\begin{figure}
	\centering
		\includegraphics[width=0.40\textwidth]{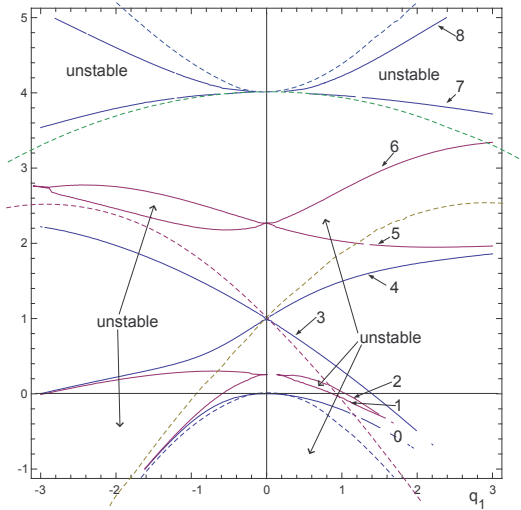}
	\caption{ Stability chart for two-elliptic CS with $\alpha^2=0.25$ shows the first nine characteristic lines with integer characteristic exponent.   The dashed lines show the characteristic lines calculated with Eq.\ref{eqw11}  for $\alpha^2=1$ (E.L.Ince chart for Mathieu's characteristic values \cite{Mclachlan:1947} ). }
	\label{fig:Spectr-0.25}
\end{figure}

\begin{figure}[htbp]
	\centering
		\includegraphics[width=0.40\textwidth]{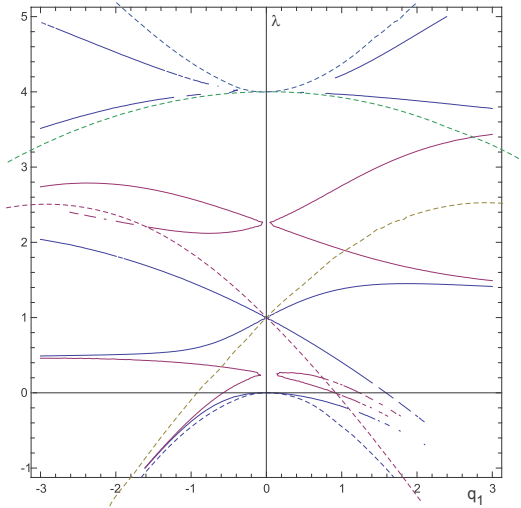}
	\caption{ Stability chart for two-elliptic CS with $\alpha^2=0.0625$. The dashed lines is Mathieu's characteristics.}
	\label{fig:Spectr_q_0_0625c}
\end{figure}
\textbf{III. Eigenvalues.}To find the characteristic curves, we need to solve transcendental equations \ref{eqw15a},\ref{eqw15b}.
It was done numerically and  Fig.\ref{fig:Spectr-0.25} presents the calculation of characteristic curves  for scale coefficient $\alpha=0.5$. Fig.\ref{fig:Spectr-0.25} demonstrates that all characteristic lines undergo the transformation and in the old regions of stability  there are several new 'peninsulas' of instability. 
The charts show some discontinuities when $q_1$ rises. This artifact is result of lack of accuracy during the solving the transcendental equations \ref{eqw15a}-\ref{eqw15b}. The real characteristic curve does not have any  discontinuity. 

There is interesting feature of charts. All of them, except the Mathieu's,  have an additional characteristic values  $0.25, 2.25, 6.25, 12.25,...$  at $q_1=0$ together with standard characteristic values $0, 1, 4, 9,...$.  These characteristic values  $0.25,...$ generated by $(n+1/2)^2$, $n=0,1,2,...$  were discovered by Meissner \cite{Meissner:1918}  and B. van der Pol, Strutt \cite{PolStrutt:1928}. As seen from whole picture of characteristics, they are  related to a discontinuity of the metric (or potential in Eq.\ref{eqw5a}, \ref{eqw6a}) at $\theta=\pi/2$, which disappears only when $\alpha=1$.    
\begin{figure}[htbp]
	\centering
		\includegraphics[width=0.40\textwidth]{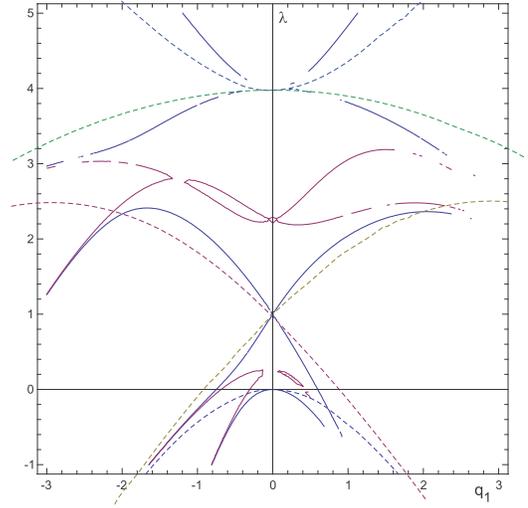}
	\caption{ Stability chart for two-elliptic CS with $\alpha^2=2.0$. The dashed lines is Mathieu's characteristics.}
	\label{fig:Spectr_q_2_0c}
\end{figure}

\begin{figure}[htbp]
	\centering
		\includegraphics[width=0.40\textwidth]{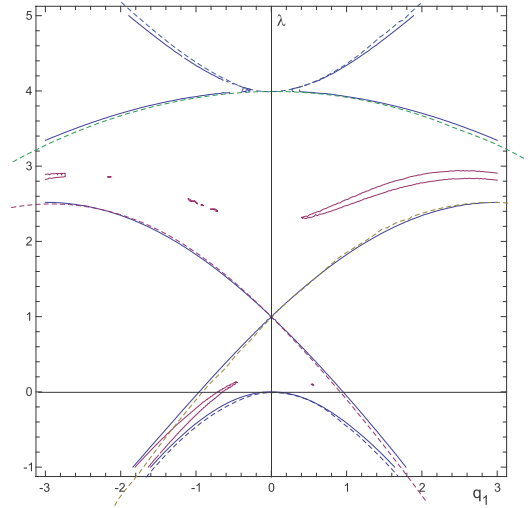}
	\caption{ Stability chart for almost symmetric two-elliptic CS with $\alpha^2=0.90625$. Interrupted lines are artifacts and the result of a failure in of calculation algorithm.  The dashed lines is Mathieu's characteristics.}
	\label{fig:Spectr_q_0_9025c_Mathieu}
\end{figure}

However, a different ratio  $\alpha$ gives a different chart of characteristic curves, Fig.\ref{fig:Spectr_q_0_0625c}-\ref{fig:Spectr_q_0_9025c_Mathieu}. The Fig.\ref{fig:Spectr_q_2_0c} was build for parameter  $\alpha >1$. When parameter $\alpha$ is almost near the unity, Fig.\ref{fig:Spectr_q_0_9025c_Mathieu}, the chart still contains the new characteristic lines up to limiting point $\alpha=1$.

Because the parameter   $\alpha$ can be varied from $0$ to $\infty$, we have an infinite number different stability charts. This gives us infinite number of new orthogonal families of coordinate systems which support the separation of variables.

\textbf{IV. Eigenfunctions.} 
In order to find the correct eigenfunctions, we need 
\begin{itemize}
\item  select the particular chart  by parameter $\alpha$; for example, the parameter $\alpha=0$ gives the family of two-elliptic systems with  degenerated right-side (semi-circles on the right side, Fig.\ref{fig:Two-elliptic}(b) shows only one such CS);    
\item
 the choice of the value $q_1$ gives a particular coordinate system from the family; the  geometry of the potential or specific boundary shape defines this choice.  

\item  'insert' the characteristic number (eigenvalue) $\lambda$  taken from intersection of $q_1$ and  characteristic curves marked by characteristic exponent: $(n=0,1,2...)$ or $(n=1/2,3/2,5/2,...)$ into the general solutions \ref{eqw7}. The solution \ref{eqw7} must be stitched on the boundaries by proper 3 coefficients from $A_1,B_1,A_2,B_2$. 

\item normalize each eigenfunction using the free coefficient.

\end{itemize}

If these can be done for all characteristic exponents, we will receive  the set of eigenfunctions 
\begin{eqnarray}
ce_{n}(\theta,  q_1, \alpha),\;\; se_{n}(\theta, q_1, \alpha)
\label{eqw20}
\end{eqnarray}
which should be complete and normalized. 

There are another alternative ways to build the  set of eigenfunctions and eigenvalues for two-elliptic CS, but described is straightforward.

\textbf{V. Conclusion.} There are very few problems for which the SE and HE 
can be solved exactly.  This is mainly due to fact that only high symmetrical boundaries and potentials allow the implementation of the powerful method of the separation of variables. Here, we suggested the two-elliptic CS which is less symmetrical than elliptic, can admit the separation of variables in HE and SE, and  includes an  infinite number of other families of CS with the same properties.  In classical dynamics, thus CS can probably unify the dispersive and regular behavior of particles.

\end{document}